\documentclass[11pt]{article}
\usepackage[margin=2.5cm]{geometry}
\usepackage{amsmath,amssymb}
\usepackage{graphicx}
\usepackage[hidelinks]{hyperref}
\usepackage{authblk}
\usepackage{booktabs}
\usepackage{xcolor}

\newif\ifmarked \markedfalse
\newcommand{\rev}[1]{\ifmarked{\color{red}#1}\else#1\fi}
\newcommand{\revon}{\ifmarked\color{red}\fi}

\newcommand{\arXivII}{arXiv:2610.30833}

\newcommand{\ud}{\mathrm{d}}

\title{Light pulse propagation through a dielectric slab.\\[0.3em]
{\revon\large I.~Retarded integral equations and their local finite-difference solution}}
\author[1]{Giancarlo Reali\thanks{reali@unipv.it}}
\affil[1]{Dipartimento di Ingegneria Industriale e dell'Informazione, Universit\`a di Pavia, via Ferrata 5a, 27100 Pavia, Italia}
\date{\rev{October 2026}}

\begin{document}
\maketitle

\begin{abstract}
We describe the propagation of a nonresonant light pulse normally incident on a thin dielectric slab using an integral-equation approach to electromagnetism. This emerges naturally from the microscopic picture of the material as made of polarizable atoms, confined within the material vacuum volume by their bonding interactions, each responding to the input radiation pulse as well as to the re-radiated pulses of the other atoms to which it is causally connected. The retarded integral equations are shown to be exactly equivalent to a local wave equation coupled to an auxiliary Lorentz-oscillator equation, which we solve numerically with a finite-difference time-domain scheme. For a slab thick enough that the front- and back-surface reflections are temporally separated, the front-surface reflectivity extracted from the simulation agrees with the Fresnel formula to within 1\%, and the back-surface echo appears at exactly the expected round-trip delay. For a slab thinner than one optical cycle, the multiply-reflected wavelets overlap and interfere, and a full space-time map of the field directly visualizes the build-up of thin-film-like interference in the time domain. The material parameters used here are illustrative rather than fitted to measured data, and the numerical scheme is verified by an explicit grid-convergence study. \rev{In a companion paper (Paper~II) the retarded integral equations are solved directly, without the local reformulation, and the two independent numerical solutions are shown to agree.}
\end{abstract}

\section{Introduction}

The reflection and transmission of light by a dielectric slab is one of the oldest problems in optics, and it is normally solved by two complementary macroscopic routes: either by imposing the Fresnel boundary conditions at each interface and adding up the resulting multiply-reflected monochromatic beams (the Airy formula for a thin film \cite{BornWolf}), or by solving Maxwell's equations directly for a medium with a prescribed refractive index $n(\omega)$. Both routes treat the medium as a continuum from the outset.

A more fundamental, microscopic route is available: the medium is modeled as a collection of polarizable atoms, each driven by the local electric field -- the superposition of the external field and the fields re-radiated by every other atom -- so that reflection, refraction and transmission emerge as a self-consistent, coherent multiple-scattering process, without ever introducing a macroscopic refractive index by hand. This program goes back at least to Ewald and Oseen (extinction-theorem), and was given a particularly transparent formulation for a single dielectric interface by Feynman (\cite{Feynman}, chap.~31); it was subsequently developed in more quantitative detail for a single interface and for a slab, in the monochromatic (continuous-wave) regime, in \cite{Reali1982,Reali1992,Reali2014}.

The availability of light sources delivering pulses only a few optical cycles long raises a natural question: how does this microscopic multiple-scattering picture behave when the driving field is not monochromatic but a short, broadband wave packet? Answering this requires promoting the continuous-wave treatment to a genuinely time-domain, causal formulation, in which retardation -- rather than a steady-state phase factor -- is the central ingredient. Beyond validating the fundamental physics in simple geometries, this spatio-temporal Finite-Difference Time-Domain and Auxiliary Differential Equation(FDTD-ADE) framework naturally opens the path toward modeling more complex physical scenarios. Because the microscopic material parameters are explicitly represented at each spatial location, the scheme can be readily extended to inhomogeneous media with spatially varying atomic density profiles $N(z)$, multi-resonant materials, or non-normal incidence, providing a versatile tool for ultrafast pulse propagation in structured optical media. 

We emphasize that the numerical method itself---the FDTD-ADE scheme for Lorentz dispersive media---is a standard technique \cite{Taflove}. The contribution of this work is not a new numerical algorithm, but rather a pedagogical reinterpretation of this standard method in terms of the retarded multiple-scattering picture. This reinterpretation makes the time-domain dynamics more transparent and connects the FDTD results directly to the microscopic extinction-theorem treatment, thereby providing a unified view of Fresnel reflection, thin-film interference, and few-cycle pulse propagation.

Section~\ref{sec:formulation} presents this formulation: the coupled retarded integral equations for the field and for the atomic polarization, obtained from a Green's-function solution of the wave equation and of the driven-oscillator equation for each polarizable atom. Section~\ref{sec:numerics} turns this formulation into a numerical scheme, by observing that the integral equations are exactly equivalent to a local system of differential equations, better suited to a FDTD discretization; the scheme is validated against the known Fresnel reflectivity and used to explore two regimes -- a slab much thicker, and one much thinner, than the pulse itself -- in which the multiple-scattering structure of the theory manifests itself very differently, from cleanly separated echo pulses to a fully interfering, thin-film-like response resolved in the time domain. Section~\ref{sec:conclusions} summarizes the results and discusses possible extensions.

\rev{This is the first of two companion papers. Here the retarded integral equations are derived and then solved through their exact local equivalent. In Paper~II~\cite{Reali2026II} the same integral equations are solved directly in the time domain, by a scheme that never passes through the local formulation, and the two numerical solutions are compared field for field; Paper~II also shows why the naive multiple-scattering (Born) series cannot be used for the parameters chosen here.}

\section{Formulation of the problem in terms of integral equations}
\label{sec:formulation}

Let us consider a dielectric plane parallel slab of (ideally) infinite extension and thickness $L$, placed with its front surface at the plane $z=0$, upon which an electromagnetic wave packet, propagating along the $z$-direction, is incident normally from the air (which henceforth we will assimilate to the vacuum). The dielectric slab is supposed to be homogeneous, isotropic and non-magnetic, and in addition a linear response to the driving electromagnetic field is assumed.

The wave equation for the electric field $E(z,t)$ (that we consider for simplicity to be a linearly polarized plane wave) is given, in Gaussian units, by
\begin{equation}
\frac{\partial^2 E(z,t)}{\partial z^2} - \frac{1}{c^2}\frac{\partial^2 E(z,t)}{\partial t^2} = \frac{4\pi}{c^2}\frac{\partial^2 P(z,t)}{\partial t^2}. \label{eq:wave}
\end{equation}
The solution of this equation provides the refracted field within the slab, as well as the reflected and transmitted electromagnetic pulses before and after the front and back surfaces of the slab, respectively.

In the microscopic picture of the dielectric, the slab is thought of as a portion of vacuum space filled with $N$ polarizable, identical atoms per unit volume, each brought into oscillation by the incident wave and by the secondary waves re-radiated by all the other oscillating atoms interacting with it. This leads to the recognition of the reflection, refraction and transmission phenomena as coherent light scattering processes (if the spectrum of the wave packet is well below the lowest absorption frequency of the dielectric, the interaction can be regarded as essentially elastic). These scattering processes are easily described in terms of integral equations that result from the passage to the continuum of the discrete picture of the atomic model and incorporate the signature of the multiple light scattering in the material \cite{GoldbergerWatson}. The integral equations also naturally emerge from a Green's function approach to the solution of Eq.~\eqref{eq:wave}.

The Green's function for the inhomogeneous problem of Eq.~\eqref{eq:wave} is the solution of the wave equation with impulsive source in space and time:
\begin{equation}
\frac{\partial^2 G(z,t)}{\partial z^2} - \frac{1}{c^2}\frac{\partial^2 G(z,t)}{\partial t^2} = \delta(z)\delta(t), \label{eq:green_def}
\end{equation}
and is given (for the one-dimensional case) by\cite{Barton} 
\begin{equation}
G(z,t) = -\frac{c}{2}\,\vartheta(t)\,\vartheta\!\left(t-\frac{|z|}{c}\right), \label{eq:green}
\end{equation}
where $\vartheta(t)=0$ for $t<0$, $=1$ for $t>0$.

With the use of this Green's function, the solution of the wave equation \eqref{eq:wave} is constructed by the standard procedure to give
\begin{align}
E(z,t) &= E_{in}(z,t) + \frac{4\pi}{c^2}\int_0^L \! \ud z' \int_{-\infty}^{\infty} \! \ud t' \, G(z-z',t-t') \frac{\partial^2 P(z',t')}{\partial t'^2} \nonumber\\
&= E_{in}(z,t) - \frac{2\pi}{c}\int_0^L \! \ud z' \int_{-\infty}^{\infty} \! \ud t' \, \vartheta(t-t')\,\vartheta\!\left(t-t'-\frac{|z-z'|}{c}\right)\frac{\partial^2 P(z',t')}{\partial t'^2} \nonumber\\
&= E_{in}(z,t) - \frac{2\pi}{c}\int_0^L \! \ud z' \int_{-\infty}^{t-|z-z'|/c} \! \ud t' \, \frac{\partial^2 P(z',t')}{\partial t'^2} \nonumber\\
&= E_{in}(z,t) - \frac{2\pi}{c}\int_0^L \! \ud z' \, \frac{\partial P(z',t-|z-z'|/c)}{\partial t}. \label{eq:Eintegral}
\end{align}
$E_{in}(z,t)$ is the incident electric field and is the solution of the homogeneous problem obtained from Eq.~\eqref{eq:wave} when the right hand side is set to zero. The electromagnetic pulse \eqref{eq:Eintegral} can be calculated at any time $t$ and at any plane within the slab for $0\le z\le L$, in front of it for $z\le 0$, and also behind it for $z\ge L$.

These internal and external fields, receiving the spatially integrated contributions of the polarization current $J_P(z,t)=\partial P(z,t)/\partial t$, itself dependent on the total internal field on the left-hand side, have to be found as self-consistent solutions of Eq.~\eqref{eq:Eintegral}, which thus has (as already noted) the typical structure of a multiple scattering problem.

In order to further proceed, we have to provide a dynamic equation for the oscillating polarized atoms, which we can rather accurately model as classical charged harmonic oscillators as far as the electromagnetic interaction with the dielectric is concerned. For monochromatic electromagnetic fields, the calculation of the polarizability $\alpha$ and of the index of refraction $n$ of the material is carried out for gases (for which $n-1\ll1$ holds) in many textbooks and also for dense materials, at a more refined level, in monographs and papers. In our case of pulse excitation, we need a time-dependent description of the polarization, which obeys the following equation:
\begin{equation}
\ddot P(z,t) + 2\gamma \dot P(z,t) + \omega_0^2 P(z,t) = \frac{Ne^2}{m}E(z,t). \label{eq:oscillator}
\end{equation}
The solution can be carried out, as in the previous case, by using the Green's function approach, where the Green's function for Eq.~\eqref{eq:oscillator} is given by\cite{Nussenzveig}
\begin{equation}
g(t) = \vartheta(t)\,e^{-\gamma t}\,\frac{\sin\!\big((\omega_0^2-\gamma^2)^{1/2}t\big)}{(\omega_0^2-\gamma^2)^{1/2}}. \label{eq:green_osc}
\end{equation}
The solution of Eq.~\eqref{eq:oscillator} then results
\begin{equation}
P(z,t) = \frac{Ne^2}{m}\int_{-\infty}^{t} g(t-t')E(z,t')\,\ud t' = \frac{Ne^2}{m}\int_0^t e^{-\gamma(t-t')}\frac{\sin\!\big(w(t-t')\big)}{w}E(z,t')\,\ud t', \label{eq:Psol}
\end{equation}
where $w=(\omega_0^2-\gamma^2)^{1/2}$ and $P(z,t=0)=0$ have been used. Indeed, we need the polarization current, which is the time derivative of the polarization. This is easily computed to be
\begin{equation}
\frac{\partial P(z,t)}{\partial t} = \frac{Ne^2}{m}\int_0^t e^{-\gamma(t-t')}\frac{w\cos\!\big(w(t-t')\big) - \gamma\sin\!\big(w(t-t')\big)}{w}E(z,t')\,\ud t'. \label{eq:Pdotsol}
\end{equation}
This is to be used in Eq.~\eqref{eq:Eintegral} evaluated at the retarded time $t-|z-z'|/c$.

The solution of the coupled system of integral equations \eqref{eq:Eintegral} and \eqref{eq:Pdotsol} is to be found by either a perturbation approach or by numerical computation, after assigning the form of the incident wave packet, of central frequency $\omega$ and width $\sigma$\rev{, travelling towards $+z$}:
\begin{equation}
\rev{E_{in}(z,t) = E_{in}(t-t_0-z/c) = e^{-\frac{(t-t_0-z/c)^2}{2\sigma^2}}\cos\!\big(\omega(t-t_0-z/c)\big),} \label{eq:pulse}
\end{equation}
where a unit incident amplitude is chosen for simplicity\rev{, and the delay $t_0$ (several times $\sigma$) places the pulse peak at the front surface $z=0$ at $t=t_0$, so that at $t=0$ the incident field is still negligible inside the slab}. For the initial conditions of the $E(z,t)$ and $P(z,t)$ fields, it is quite natural to take them as follows:
\begin{equation}
\rev{E(z,t{=}0)=E_{in}(z,t{=}0),\qquad \partial_tE(z,t{=}0)=\partial_tE_{in}(z,t{=}0),\qquad} P(z,t{=}0)=0,\qquad \dot P(z,t{=}0)=0. \label{eq:ic}
\end{equation}
The solutions will then provide the internal field $E(z,t)$ at any plane within the dielectric slab. In turn, the reflected field $E_r$ for $z\le 0$, and the transmitted field $E_t$ for $z\ge L$ can be computed. In particular, we will give the results for these two fields at the front and back interfaces, $E_r(z{=}0,t)$ and $E_t(z{=}L,t)$:
\begin{align}
E_r(z{=}0,t) &= -\frac{2\pi}{c}\int_0^L \! \ud z' \, \frac{\partial P(z',t-z'/c)}{\partial t}, \label{eq:Er}\\
E_t(z{=}L,t) &= E_{in}(L,t) - \frac{2\pi}{c}\int_0^L \! \ud z' \, \frac{\partial P(z',t-(L-z')/c)}{\partial t}. \label{eq:Et}
\end{align}

\subsection{Remark on the initial condition \eqref{eq:ic}.} 

It is worth stressing that Eq.~\eqref{eq:ic} is a statement about the value of the fields \emph{at $t=0$}, not a boundary condition to be enforced at $z=0$ for all subsequent times. Since $P(z,0)=0$ everywhere, Eq.~\eqref{eq:Er} gives $E_r(0,0)=0$ identically: only the incident field can be present anywhere in space at the initial instant, simply because the reflected field has not yet had time to form -- this is a direct statement of causality, exact rather than approximate. For $t>0$ the front-surface field is, in general, $E(0,t)=E_{in}(0,t)+E_r(0,t)$ with $E_r$ generally different from zero, as prescribed self-consistently by the coupled system \eqref{eq:Eintegral} and \eqref{eq:Pdotsol}. This distinction matters in practice: a numerical scheme that mistakenly clamps $E(0,t)=E_{in}(0,t)$ for \emph{every} $t$ would suppress the very reflected field one wants to compute. Section~\ref{sec:numerics} describes a source-injection scheme designed specifically to avoid this pitfall.

\subsection{From discrete dipoles to continuous polarization}

Before turning to the numerical solution, it is useful to clarify the relationship between the microscopic picture of discrete polarizable atoms and the continuous polarization field \(P(z,t)\) used in Eq. \eqref{eq:oscillator}. This transition involves two key approximations.

First, the discrete atomic polarizability \(\alpha\) is related to the refractive index \(n(\omega)\) through the Clausius--Mossotti relation (in Gaussian units):

\begin{equation}
\frac{n^2(\omega) - 1}{n^2(\omega) + 2} = \frac{4\pi}{3} N \alpha(\omega),
\label{eq:clausius}
\end{equation}
which reduces, in the dilute-gas limit \(n \approx 1\), to the simpler form

\begin{equation}
n^2(\omega) - 1 \approx 4\pi N \alpha(\omega).
\label{eq:dilute}
\end{equation}

For the illustrative parameters used here (Table~\rev{\ref{tab:params}}), the target refractive index \(n = 1.45\) is a typical dense-medium value, so the dilute-gas approximation \eqref{eq:dilute} is a simplification of convenience rather than a quantitatively justified limit; a fully self-consistent local-field treatment \eqref{eq:clausius} would be needed for accuracy. We use the dilute-gas expression \eqref{eq:dilute} for simplicity, as it is sufficient to reproduce the qualitative reflection/transmission dynamics. 

Second, the transition from discrete dipoles to the continuous polarization field \(P(z,t)\) requires coarse-graining over a volume containing many atoms. This is justified when the atomic spacing is much smaller than both the optical wavelength and the spatial variations of the field. For the parameters used here (a typical atomic density \(N \sim 10^{22}\) cm\(^{-3}\) for a solid), the interatomic spacing is of order \rev{$N^{-1/3}\approx0.5$~nm}, while the optical wavelength is 800 nm; the coarse-graining is therefore well justified.

It is important to stress, however, that the numerical scheme we implement solves the \textit{continuous} Maxwell--Lorentz equations \eqref{eq:wave} and \eqref{eq:oscillator}, not the discrete retarded dipole equations. The integral formulation of Sec.~\ref{sec:formulation} provides a causal, multiple-scattering \textit{interpretation} of these equations, but it is not a new numerical method for simulating discrete atoms. A genuine discrete-dipole simulation would require solving the coupled equations for each dipole individually, including the retarded Green's function between all pairs---a different and computationally far more demanding problem. Our goal here is pedagogical: to show that the standard local-field FDTD calculation can be \textit{understood} as a multiple-scattering process, and that this understanding illuminates the time-domain build-up of reflection and interference.

\section{Numerical solution}
\label{sec:numerics}

The solution of the coupled system of integral equations \eqref{eq:Eintegral} and \eqref{eq:Pdotsol} can in principle be approached perturbatively, order by order in the multiple-scattering strength, or by direct numerical computation. In the case of strongly correlated, coherently interacting dipoles -- such as the thin-slab regime considered in Sec.~\ref{sec:thin} -- \rev{the perturbative (Born) series is not a usable route: as shown in Paper~II~\cite{Reali2026II}, its time-domain iteration diverges for the parameters used here,} and the problem is \rev{instead} solved numerically, which is the route we take in the following.

\subsection{Equivalence with a local differential formulation}
\label{sec:equivalence}

Equation \eqref{eq:Eintegral} is, by construction, nothing but the formal solution of the local wave equation \eqref{eq:wave} obtained via the retarded Green's function \eqref{eq:green}; likewise Eq.~\eqref{eq:Pdotsol} is the formal solution of the local oscillator equation \eqref{eq:oscillator}. It follows that the pair of \emph{non-local, retarded, integral} equations \eqref{eq:Eintegral} and \eqref{eq:Pdotsol} is exactly equivalent to the pair of \emph{local} differential equations \eqref{eq:wave} and \eqref{eq:oscillator}, supplemented with the initial data \eqref{eq:ic}. Rather than discretizing the convolution integral directly -- which requires special care of the ``self-interaction'' term where the retardation $|z-z'|/c\to0$, i.e. exactly where a finite-difference grid cell scatters onto itself -- we solve numerically the equivalent local system. \rev{(The direct discretization, with an algebraic treatment of this zero-delay term, is the subject of Paper~II~\cite{Reali2026II}.)} This is the standard strategy used in FDTD electromagnetics, where Eq.~\eqref{eq:oscillator} is treated as an ADE for a Lorentz-model polarization coupled to the wave equation \cite{Taflove}. The two formulations are mathematically equivalent; the local one is simply free of the \rev{zero-delay self-interaction term, which in a direct discretization makes the update implicit,} and \rev{is} considerably easier to discretize to second-order accuracy. On the other end, the integral formulation recasts the local-field equations as a self-consistent multiple-scattering problem, which is pedagogically valuable for understanding the time-domain build-up of the reflected and transmitted fields.

\subsection{Material and pulse parameters}
\label{sec:parameters}

We take, as a reference material, a single-resonance illustrative dielectric with a dominant electronic resonance in the ultraviolet, consistently with the assumption (Sec.~\ref{sec:formulation}) that the incident visible pulse is spectrally far from the lowest absorption line, so that the imaginary part of the response can be kept small without qualitatively affecting the reflection/transmission dynamics. The resonance parameters $\omega_0$, $\gamma$ and the coupling constant $Ne^2/m$ are fixed as follows: (i) the resonance wavelength is placed in the deep UV, $\lambda_0=120\,\mathrm{nm}$; (ii) the damping is parametrized through a quality factor (\cite{Feynman}, chap.~23) $Q=\omega_0/2\gamma=50$, small enough to be numerically well resolved and consistent with a narrow, weak UV absorption line; (iii) the coupling constant $Ne^2/m$ is fixed by requiring that the \emph{undamped} Lorentz formula
\begin{equation}
n^2(\omega) - 1 = \frac{4\pi\, Ne^2/m}{\omega_0^2-\omega^2}
\label{eq:sellmeier}
\end{equation}
reproduce the target refractive index $n=1.45$ at the pulse carrier frequency $\omega$. From the same (undamped) dispersion relation we also compute, by a centered finite difference, the group index at the carrier frequency,
\begin{equation}
n_g(\omega) = n(\omega) + \omega\,\frac{dn}{d\omega}\bigg|_{\omega},
\label{eq:groupindex}
\end{equation}
which governs the transit time of the pulse \emph{envelope} through the slab and will be checked directly against the simulation in Sec.~\ref{sec:groupindex}. The incident pulse, Eq.~\eqref{eq:pulse}, is centered at the visible wavelength $\lambda_c=2\pi c/\omega = 800\,\mathrm{nm}$ ($T_{cycle}=2.67\,\mathrm{fs}$) with a Gaussian envelope \rev{whose field FWHM is $5.9\,\mathrm{fs}$ ($2.2$ optical cycles; $\sigma=5.9\,\mathrm{fs}/(2\sqrt{2\ln2})=2.5\,\mathrm{fs}$ in Eq.~\eqref{eq:pulse}), i.e.\ an intensity FWHM of $4.2\,\mathrm{fs}$}: a few-cycle pulse. Table~\ref{tab:params} summarizes these common parameters; the slab thickness $L$ is instead varied between the two regimes discussed below.
\begin{table}[!htbp]
\centering
\begin{tabular}{@{}ll@{}}
\toprule
Quantity & Value \\
\midrule
Carrier wavelength $\lambda_c$ & 800 nm \\
Optical cycle $T_{cycle}$ & 2.67 fs \\
\rev{Pulse FWHM, field envelope} & 5.9 fs (2.2 cycles) \\
\rev{Pulse FWHM, intensity} & \rev{4.2 fs ($\approx1.6$ cycles)} \\
UV resonance wavelength $\lambda_0$ & 120 nm ($\omega_0/\omega \approx 6.7$) \\
Quality factor $Q=\omega_0/2\gamma$ & 50 \\
Target refractive index $n(\omega)$ & 1.45 \\
Group index $n_g(\omega)=n+\omega\,dn/d\omega$ & 1.4675 \\
Group velocity $v_g=c/n_g$ & $0.6814\,c$ \\
Fresnel reflectivity, single interface, $R_F=\left(\frac{n-1}{n+1}\right)^2$ & 0.0337 \\
\bottomrule
\end{tabular}
\caption{Material and pulse parameters common to both simulated regimes.}
\label{tab:params}
\end{table}
We stress that this is a \emph{toy calculation}: $\lambda_0$ and $Q$ are chosen to place the resonance safely in the UV and to keep the absorption line numerically well resolved, not to fit the measured dispersion of any specific glass. A quantitative comparison with real materials -- e.g.\ a multi-term Sellmeier fit to fused silica, or a measured absorption line shape -- is deferred to future work; here the goal is to test the numerical scheme and to illustrate the two qualitative regimes (separated vs.\ interfering echoes) discussed in Sec.~\ref{sec:results}.

\subsection{Numerical scheme}

Equations \eqref{eq:wave} and \eqref{eq:oscillator} are discretized on a uniform space-time grid with a standard second-order, explicit, leapfrog scheme (grid spacing and time step are chosen well within the Courant stability limit for each regime, see Sec.~\ref{sec:results}). At each time step the polarization is first advanced using a central-difference discretization of Eq.~\eqref{eq:oscillator},
\begin{equation}
P^{n+1}_i \left(\frac{1}{\Delta t^2}+\frac{\gamma}{\Delta t}\right) = \frac{Ne^2}{m}E^n_i - \omega_0^2 P^n_i + \frac{2P^n_i}{\Delta t^2} + P^{n-1}_i\left(\frac{\gamma}{\Delta t}-\frac{1}{\Delta t^2}\right),
\label{eq:ade}
\end{equation}
using only the already-known field $E^n_i$ (an important sign check: the coefficient of $P^{n-1}_i$ must be $\gamma/\Delta t-1/\Delta t^2$, \emph{not} its opposite -- the wrong sign turns the scheme into an exponentially unstable one, as we verified numerically). The field is then advanced with the leapfrog update of the wave equation \eqref{eq:wave},
\begin{equation}
E^{n+1}_i = 2E^n_i - E^{n-1}_i + \left(\frac{c\,\Delta t}{\Delta z}\right)^2\!\big(E^n_{i+1}-2E^n_i+E^n_{i-1}\big) - 4\pi\big(P^{n+1}_i-2P^n_i+P^{n-1}_i\big).
\label{eq:leapfrog}
\end{equation}

Consistently with the discussion following Eq.~\eqref{eq:ic}, the incident pulse is \emph{not} imposed as a value clamped at $z=0$ for all $t$. Instead we use a standard total-field/scattered-field (TF/SF) one-way source injection: the known, exact, free-space solution $E_{in}(z,t)$ of Eq.~\eqref{eq:pulse} is injected at a single plane $z_{src}$, well inside the vacuum region and away from both the slab and the domain boundaries, through the interface correction terms
\begin{equation}
E^{n+1}_{src} \mathrel{+}= \left(\frac{c\Delta t}{\Delta z}\right)^2 E_{in}(z_{src}-\Delta z,t_n), \qquad
E^{n+1}_{src-1} \mathrel{-}= \left(\frac{c\Delta t}{\Delta z}\right)^2 E_{in}(z_{src},t_n),
\label{eq:tfsf}
\end{equation}
applied after the ordinary update \eqref{eq:leapfrog}. This injects a purely right-going wave: the region to the left of $z_{src}$ is left as a pure scattered-field zone which, by construction, contains no incident-field component and is therefore identically zero until a genuine reflection physically arrives -- precisely realizing, numerically, the causal content of Eq.~\eqref{eq:ic} discussed above. First-order Mur absorbing boundary conditions are applied at both ends of the computational domain, so that no field is ever externally prescribed other than through the TF/SF injection \eqref{eq:tfsf}. The reflected field $E_r(0,t)$ of Eq.~\eqref{eq:Er} is read directly off a probe placed in the scattered-field region (a retarded copy of it, since vacuum propagation only delays the waveform without reshaping it); the transmitted field $E_t(L,t)$ of Eq.~\eqref{eq:Et} is read directly off a probe placed beyond the slab, where the only field present is, by construction, the physical total field.

\subsection{Results: two regimes}
\label{sec:results}

Two slab thicknesses are simulated, with everything else (material, pulse, numerical scheme) held fixed as in Sec.~\ref{sec:parameters}: a ``thick'' slab, $L=3\,\mu\mathrm{m}$ ($\Delta z=15\,\mathrm{nm}$, $\Delta t\approx20\,\mathrm{as}$), for which the round-trip time $2nL/c=29.0\,\mathrm{fs}$ is much longer than the pulse duration, so that successive reflections are temporally well separated; and a ``thin'' slab, $L=150\,\mathrm{nm}$ ($\Delta z=5\,\mathrm{nm}$, $\Delta t\approx6.7\,\mathrm{as}$), for which the round-trip time $2nL/c=1.45\,\mathrm{fs}$ is shorter than a single optical cycle, so that the reflected wavelets from the front and back surfaces necessarily overlap and interfere.

\subsubsection{Thick slab: separated echoes and a Fresnel-reflectivity check}

\begin{figure}[!htbp]
\centering
\includegraphics[width=0.8\textwidth]{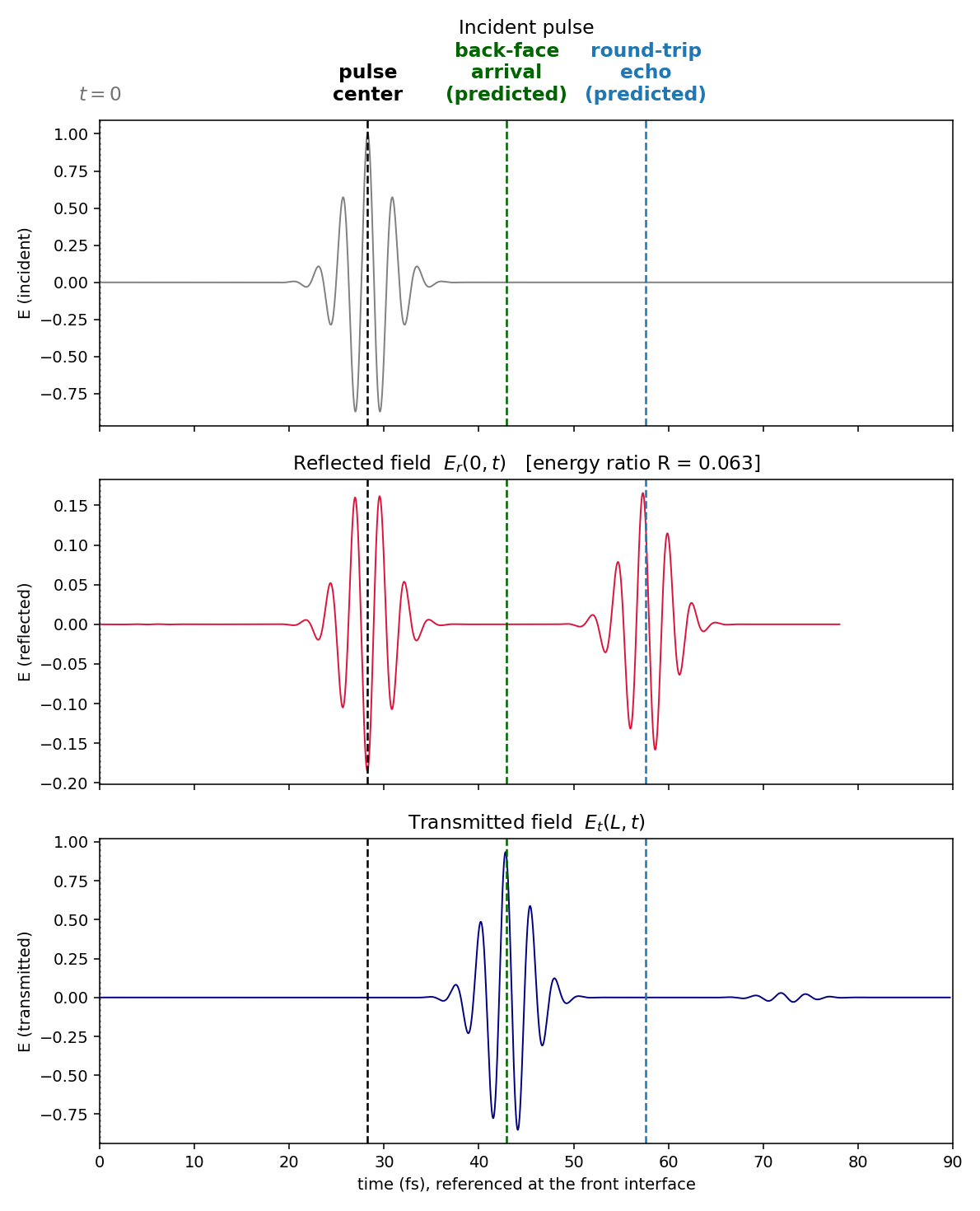}
\caption{Thick slab ($L=3\,\mu\mathrm{m}$), all three panels on a common time axis referenced at the front interface. Top: incident pulse at the slab front. Middle: reflected field $E_r(0,t)$; the two wavepackets are the front-surface reflection and the back-surface echo, separated by the round-trip time $2nL/c=29\,\mathrm{fs}$. Bottom: transmitted field $E_t(L,t)$, delayed and weakly reshaped relative to the incident pulse, with a small secondary ripple due to the internally-reflected replica. Vertical dashed lines, common to all three panels, mark: the time origin $t=0$ (grey); the incident-pulse center, i.e.\ its peak arrival time at the front interface (black); the \emph{predicted} one-way group-delay arrival at the back interface, $t_{\mathrm{peak}}+n_gL/c$ (green); and the \emph{predicted} round-trip echo time, $t_{\mathrm{peak}}+2n_gL/c$ (blue) -- see Sec.~\ref{sec:groupindex} for the group index $n_g$ used and for a quantitative comparison with the measured delay.}
\label{fig:thick}
\end{figure}

Figure~\ref{fig:thick} shows the incident pulse together with the reflected and transmitted fields for the thick slab. Three features, all consequences of the multiple-scattering structure of Eq.~\eqref{eq:Eintegral}, are apparent.

First, the front-surface reflection tracks the incident pulse in time and has a peak amplitude of $\approx0.18$ in units of the unit-amplitude incident pulse, in good agreement with the Fresnel amplitude reflection coefficient $r=(n-1)/(n+1)\approx0.184$ for $n=1.45$. Integrating the reflected energy over only this first feature gives a numerical reflectivity of $0.034$, matching the single-interface Fresnel value $R_F=0.0337$ of Table~\ref{tab:params} to within $1\%$ -- a validation of the numerical scheme (source injection, ADE, absorbing boundaries) against the known continuous-wave limit.

Second, a delayed feature appears in $E_r(0,t)$ at $\Delta t\approx29\,\mathrm{fs}$ after the first, in essentially exact agreement with the slab round-trip time. This is the pulse that reflects off the back surface of the slab and re-emerges through the front surface: a direct, time-domain manifestation of the multiple internal reflections that the microscopic multiple-scattering picture of Sec.~\ref{sec:formulation} is built to describe self-consistently, and that a naive single-surface Fresnel treatment would entirely miss. A correspondingly small secondary ripple, delayed by the same round-trip time, is visible after the main peak of $E_t(L,t)$.

Third, the transmitted pulse $E_t(L,t)$ arrives later than a pulse freely propagating in vacuum over the same distance would, consistently with the reduced group velocity $v_g=c/n_g<c/n$ inside the slab (Table~\ref{tab:params}), and shows a small amount of reshaping (asymmetric ringing at the trailing edge) -- a signature of the frequency-dependent (dispersive) response of the Lorentz medium acting on the finite bandwidth of a few-cycle pulse. Section~\ref{sec:groupindex} quantifies this delay.

An additional global consistency check is provided by energy conservation. For a weakly dissipative medium (quality factor \(Q = 50\) corresponds to a small but nonzero absorption), the time-integrated Poynting fluxes satisfy

\begin{equation}
R + T + A = 1,
\label{eq:energy}
\end{equation}
where \(R\) and \(T\) are the reflected and transmitted energy fractions, and \(A\) is the fraction absorbed in the slab. $R$ and $T$ are obtained from the time-integrated flux $(c/4\pi)\!\int\! E^2\,dt$ of the (purely traveling-wave) reflected and transmitted fields, normalized to the same integral for the incident pulse; $A$ is obtained independently, from the volume- and time-integrated dissipation-rate density $(2\gamma/\kappa)(\partial P/\partial t)^2$ of the driven, damped oscillator (with $\kappa\equiv Ne^2/m$), integrated over the slab and normalized the same way -- an identity that follows directly from the Poynting theorem for the coupled Maxwell--Lorentz system. Evaluating these three integrals independently gives \rev{$R=0.0626$, $T=0.8826$, $A=0.0542$ (sum $0.9994$)} for the thick slab, and $R=0.120$, $T=0.878$, $A=0.003$ (sum $1.000$) for the thin slab; in both cases the deviation of $R+T+A$ from unity is below $0.1\%$, consistent with the residual discretization error quantified in Sec.~\ref{sec:convergence}. The absorbed fraction is markedly larger for the thick slab, as expected since the pulse travels a much longer path inside the (weakly) lossy medium ($3\,\mu\mathrm{m}$ vs.\ $150\,\mathrm{nm}$). This confirms the internal consistency of the FDTD-ADE scheme and the correctness of the absorbing boundary conditions.

\subsubsection{Group-index check}
\label{sec:groupindex}

Equation~\eqref{eq:groupindex} gives, for the parameters of Table~\ref{tab:params}, $n_g=1.4675$ (to be compared with $n=1.45$), corresponding to a group velocity $v_g=c/n_g=0.6814\,c$. The predicted one-way group delay through the $L=3\,\mu\mathrm{m}$ thick slab is then $n_gL/c=14.685\,\mathrm{fs}$. Measuring directly, in the simulation, the time interval between the incident-pulse peak at the front interface and the transmitted-pulse peak at the back interface (the black and green markers of Fig.~\ref{fig:thick}, respectively) gives $14.560\,\mathrm{fs}$, a difference of $0.85\%$ from the dispersion-relation prediction. Given that the pulse has a finite bandwidth (so that $n_g$ evaluated at the single carrier frequency is itself only an approximation to the true envelope transit time) and that a small residual discretization error is expected at the resolution used (Sec.~\ref{sec:convergence}), this level of agreement is a further, independent validation of the numerical scheme -- and a direct illustration of why $n_g$, rather than $n$, is the relevant quantity for pulse-propagation timing.

\subsubsection{Thin slab: interference between reflected wavelets}
\label{sec:thin}

\begin{figure}[!htbp]
\centering
\includegraphics[width=0.8\textwidth]{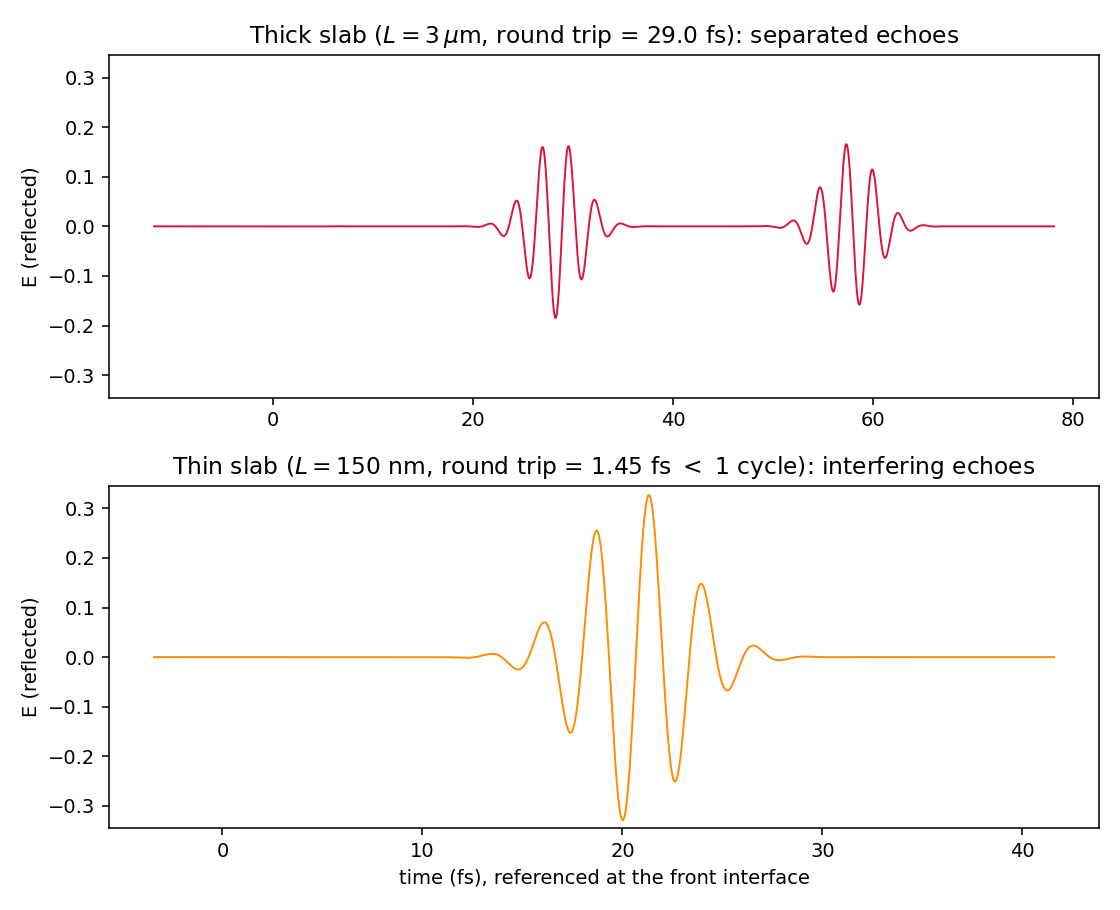}
\caption{Reflected field $E_r(0,t)$ for the thick slab of Fig.~\ref{fig:thick} (top) compared with the thin slab, $L=150\,\mathrm{nm}$ (bottom), \emph{both panels plotted on the same vertical scale}: the field is normalized throughout this paper so that the incident pulse has unit peak amplitude, so the two curves are directly comparable without any additional rescaling. In the thin-slab case the round trip ($1.45\,\mathrm{fs}$) is shorter than one optical cycle ($2.67\,\mathrm{fs}$): the front- and back-surface reflections merge into a single, reshaped wavepacket rather than two separated echoes, with a peak amplitude ($\approx0.32$) and an integrated reflected energy ($R\approx0.12$) markedly larger than for either a single interface ($R_F=0.034$) or the incoherent sum of two independent reflections -- a signature of constructive interference between the two reflected wavelets.}
\label{fig:thin_vs_thick}
\end{figure}

When the slab is thin enough that $2nL/c$ is shorter than the pulse duration -- and, for the parameters of Fig.~\ref{fig:thin_vs_thick}, even shorter than a single optical cycle -- the front- and back-surface reflections can no longer be resolved as separate wavepackets: they overlap in time and interfere, exactly as in the classical treatment of thin-film interference (the Airy formula, \cite{BornWolf}), but here obtained self-consistently, in the time domain, from the microscopic multiple-scattering equations, and for a genuinely broadband few-cycle pulse rather than a monochromatic wave.

Figure~\ref{fig:thin_vs_thick} contrasts the two regimes directly. Rather than the two clean echo pulses of Fig.~\ref{fig:thick}, the thin-slab reflected field is a single, asymmetric wavepacket with a peak amplitude ($\approx0.32$) and integrated energy ($R\approx0.12$) considerably larger than the single-interface Fresnel value, and larger even than the incoherent sum of two independent reflections ($2R_F\approx0.067$): at the particular optical thickness chosen here, the two reflected wavelets add constructively over most of their overlap, in direct analogy with a thin film sitting near a constructive-interference (bright) fringe of the Airy formula -- except that here the ``fringe'' is resolved pulse by pulse in time, rather than swept in wavelength or thickness at a single, monochromatic frequency. \rev{Indeed, the optical thickness $nL=217.5$~nm is close to a quarter of the carrier wavelength ($nL/\lambda_c=0.27$), the condition for maximum reflectance of a single layer, $R_{\max}=[(n^2-1)/(n^2+1)]^2=0.126$; the monochromatic Airy reflectance at the carrier, $0.124$, is close to the value $R\approx0.12$ found for the broadband pulse.}

\begin{figure}[!htbp]
\centering
\includegraphics[width=0.78\textwidth]{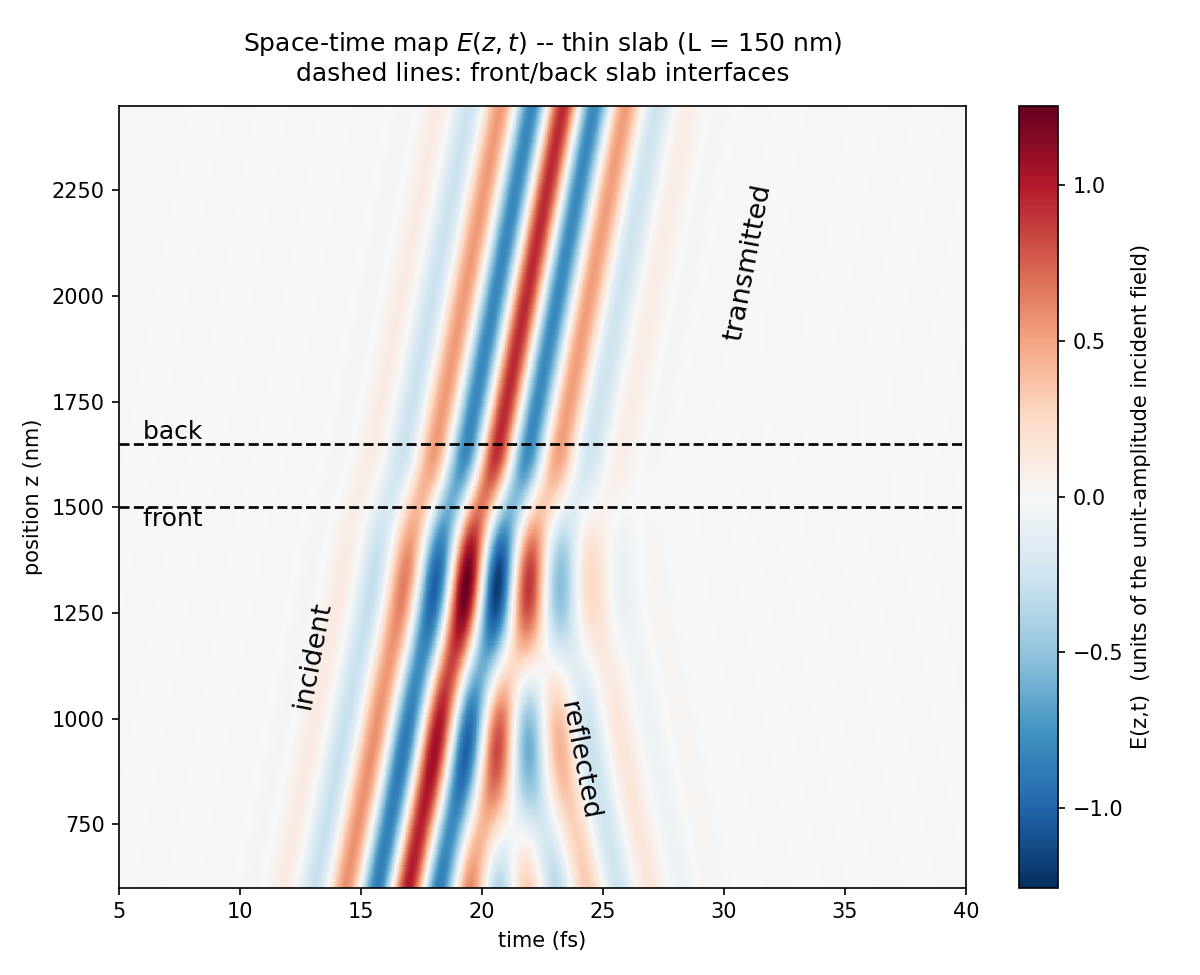}
\caption{Space-time map of the raw oscillating field $E(z,t)$ (not its envelope), in units of the unit-amplitude incident field, for the thin slab ($L=150\,\mathrm{nm}$). Dashed horizontal lines mark the front and back slab interfaces. The incident wavepacket (propagating up and to the right at speed $c$) partially reflects at the front interface (propagating down and to the right, i.e.\ backward in $z$) and partially transmits through the slab, continuing beyond the back interface with a slight group delay. The cusp-shaped region immediately below the front interface, where incident and reflected fields spatially overlap, contains the spatial interference fringes discussed in the text; see Fig.~\ref{fig:envelope} for the corresponding envelope-only view.}
\label{fig:spacetime}
\end{figure}

The full spatio-temporal build-up of this interference is shown in Fig.~\ref{fig:spacetime}, a map of the raw carrier field $E(z,t)$ with the front and back slab interfaces marked by dashed lines. The incident wavepacket (labelled ``incident'') is seen propagating to the right at speed $c$ until it reaches the front interface; a reflected wavepacket (labelled ``reflected'') then emerges, propagating to the left, while a transmitted wavepacket (labelled ``transmitted'') continues to the right beyond the back interface, delayed by its slower group velocity inside the slab.

\begin{figure}[!htbp]
\centering
\includegraphics[width=0.78\textwidth]{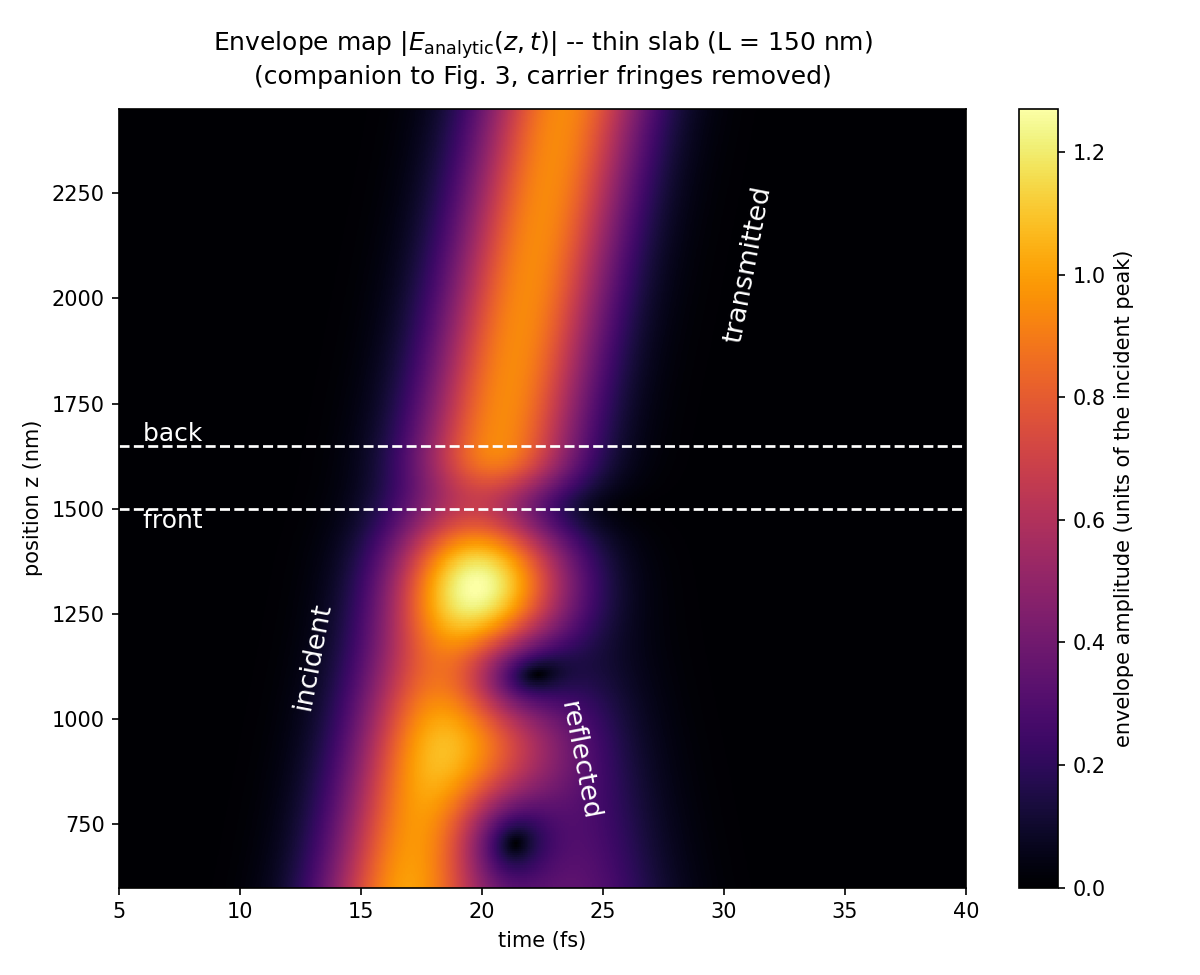}
\caption{Envelope map $|E_{\mathrm{analytic}}(z,t)|$ (Hilbert transform in time at each $z$) for the same simulation as Fig.~\ref{fig:spacetime}, with the carrier fringes removed. The bright spot straddling the front interface is the constructive-interference enhancement of Fig.~\ref{fig:thin_vs_thick}; the darker regions immediately adjacent to it correspond to partially destructive interference between the directly-reflected wavelet and the one that has completed an internal round trip.}
\label{fig:envelope}
\end{figure}

The cusp-like structure below the front interface deserves a more careful physical reading, precisely because Fig.~\ref{fig:spacetime} shows the raw carrier oscillation rather than its envelope. In that overlap region, two wavetrains of essentially the same optical frequency propagate in opposite directions: the tail of the still-arriving incident pulse (moving toward $+z$) and the already-generated front-surface reflection (moving toward $-z$). Two counter-propagating waves of equal frequency interfere in space with a fringe spacing of $\lambda_c/2\approx400\,\mathrm{nm}$ (a static, ``standing-wave-like'' pattern, not a beat note in time), so the fine alternating red/blue stripes seen inside the cusp are these spatial interference fringes, not a feature of either wavepacket individually. The cusp \emph{shape} itself is purely kinematic: it is bounded by the light-cone-like trajectories of the leading and trailing edges of the two counter-propagating envelopes, so the region of temporal overlap -- where interference is possible at all -- shrinks the further one looks in $z$ away from the interface, tracing out the triangular/cusp outline. What Fig.~\ref{fig:spacetime} does \emph{not} show directly is where this interference is constructive or destructive on the envelope scale: for that, the envelope map of Fig.~\ref{fig:envelope} is the more appropriate, and more broadly accessible, visualization.

As a pedagogical companion to Fig.~\ref{fig:spacetime}, Fig.~\ref{fig:envelope} shows the envelope $|E_{\mathrm{analytic}}(z,t)|$ of the same field, obtained from a Hilbert transform in time at each fixed $z$, which removes the fast carrier oscillations and displays directly how the pulse \emph{energy} (rather than the instantaneous field) splits at the interface. The reflected and transmitted wavepacket envelopes are now visible as smooth, single-lobed features, and -- most usefully -- the envelope map reveals directly, as a bright spot straddling the front interface, the constructive-interference enhancement already quantified in Fig.~\ref{fig:thin_vs_thick} ($R\approx0.12$ vs.\ $R_F=0.034$ for a single interface), flanked by two darker, partially destructive-interference regions immediately below and to either side of it. This is the phase relationship between the directly-reflected front-surface wavelet and the wavelet having made one internal round trip, made visible on the envelope scale rather than inferred from the carrier fringes of Fig.~\ref{fig:spacetime}.

\begin{figure}[!htbp]
\centering
\includegraphics[width=0.75\textwidth]{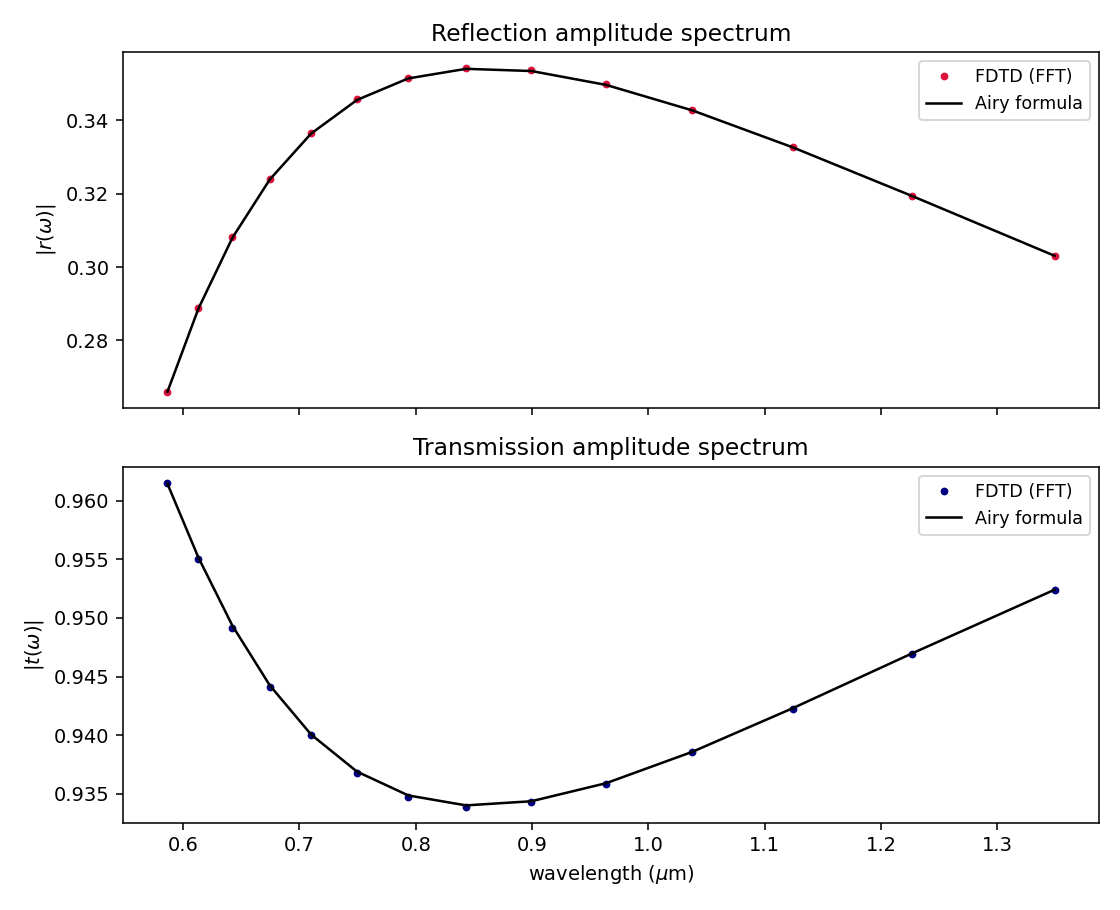}
\caption{Reflection and transmission amplitude spectra $|r(\omega)|$, $|t(\omega)|$ for the thin slab, extracted from the FDTD fields by Fourier transform (points), compared with the analytic Airy formula \eqref{eq:airy_freq} evaluated with the same complex Lorentz index used in the simulation (solid line). The spectral-weighted RMS relative deviation is $0.05\%$ for $|r(\omega)|$ and $0.01\%$ for $|t(\omega)|$.}
\label{fig:airy}
\end{figure}

As a further consistency check, we compare the FDTD results with the frequency-domain Airy prediction. For a linear, time-invariant system, the Fourier transform of the transmitted (or reflected) field is related to the incident spectrum by

\begin{equation}
\tilde{E}_{t}(\omega) = t_{Airy}(\omega)\tilde{E}_{in}(\omega), \qquad \tilde{E}_{r}(\omega) = r_{Airy}(\omega)\tilde{E}_{in}(\omega),
\label{eq:airy_freq}
\end{equation}
where \(t_{Airy}(\omega)\) and \(r_{Airy}(\omega)\) are the standard Airy transmission and reflection coefficients for a thin film \cite{BornWolf}, evaluated with the \emph{same} complex, frequency-dependent Lorentz index $\tilde n(\omega)=\sqrt{1+4\pi\kappa/(\omega_0^2-\omega^2-2i\gamma\omega)}$ used throughout the FDTD simulation (not just its value $n=1.45$ at the carrier frequency). We compute \(\tilde{E}_{in}(\omega)\) analytically from Eq.~\eqref{eq:pulse}, and \(\tilde{E}_{r}(\omega)\), \(\tilde E_t(\omega)\) numerically from the FDTD reflected and transmitted fields via a fast Fourier transform, restricting the comparison to the part of the spectrum containing at least $5\%$ of the incident spectral peak. Figure~\ref{fig:airy} shows the resulting amplitude spectra $|r(\omega)|$ and $|t(\omega)|$: the FDTD points fall on the analytic Airy curves to within a spectral-weighted RMS relative deviation of $0.05\%$ for $|r(\omega)|$ and $0.01\%$ for $|t(\omega)|$, confirming that the time-domain FDTD results are fully consistent with the frequency-domain Airy treatment, as expected from the linearity of the Maxwell--Lorentz system.

\subsection{Numerical convergence}
\label{sec:convergence}

Since the thin-slab case of Sec.~\ref{sec:thin} is numerically the more demanding of the two (shortest length and time scales, strongest sensitivity to how well the sub-cycle interference is resolved), we use it to check the convergence of the scheme, refining the grid spacing $\Delta z$ by successive factors of 2 (with $\Delta t = 0.4\,\Delta z/c$ kept proportional, i.e.\ at fixed Courant number) while keeping the physical domain and all material/pulse parameters fixed. Table~\ref{tab:convergence} reports two diagnostics -- the integrated reflected-energy ratio $R$ and the peak reflected-field amplitude -- at four resolutions, from $\Delta z=20\,\mathrm{nm}$ (coarser than the value $\Delta z=5\,\mathrm{nm}$ used to produce Figs.~\ref{fig:thin_vs_thick}--\ref{fig:envelope}) down to $\Delta z=2.5\,\mathrm{nm}$.

\begin{table}[!htbp]
\centering
\begin{tabular}{@{}lcccc@{}}
\toprule
$\Delta z$ (nm) & $\Delta t$ (as) & $R$ & peak $|E_r|$ \\
\midrule
20.0 & 26.68 & 0.11603 & 0.32186 \\
10.0 & 13.34 & 0.11984 & 0.32915 \\
5.0  & 6.67  & 0.11948 & 0.32879 \\
2.5  & 3.34  & 0.11939 & 0.32870 \\
\bottomrule
\end{tabular}
\caption{Grid-convergence study for the thin-slab reflected field (integrated energy ratio $R$ and peak amplitude), at fixed Courant number $c\Delta t/\Delta z=0.4$. The resolution used elsewhere in this paper for the thin slab is $\Delta z=5\,\mathrm{nm}$.}
\label{tab:convergence}
\end{table}

Inspecting the sequence in Table~\ref{tab:convergence}, $R$ is \emph{not} monotonic across all four levels: it increases from $\Delta z=20$ to $10\,\mathrm{nm}$ ($0.11603\to0.11984$) before decreasing for the two finer steps ($0.11984\to0.11948\to0.11939$); the peak amplitude shows the same pattern. This indicates that $\Delta z=20\,\mathrm{nm}$ lies outside the asymptotic (small-$\Delta z$) convergence regime of the scheme, so a Richardson order estimate built from the three coarsest levels is not meaningful (we find nominal $p\approx3.4$ and $p\approx4.4$ for $R$ and the peak amplitude respectively from that triplet, values with no direct interpretation given the sign change). Restricting the estimate to the three \emph{finest} levels ($\Delta z=10,5,2.5\,\mathrm{nm}$), where the sequence decreases monotonically and consistently for both diagnostics, gives an observed order $p\approx2.09$ for $R$ and $p\approx2.00$ for the peak amplitude, consistent with the formal second-order accuracy of the leapfrog scheme \eqref{eq:ade}--\eqref{eq:leapfrog}. The relative change between the two finest levels is below $0.1\%$ for $R$ and below $0.03\%$ for the peak amplitude, so the $\Delta z=5\,\mathrm{nm}$ resolution used to generate Figs.~\ref{fig:thin_vs_thick}--\ref{fig:envelope} is already converged to well within plotting accuracy, even though the coarsest level tested is not yet in the asymptotic regime. An analogous check on the thick-slab front-surface reflectivity of Fig.~\ref{fig:thick} (not shown in full) gives $R_{front}=0.0352,\,0.0341,\,0.0338$ at $\Delta z=30,15,7.5\,\mathrm{nm}$ respectively -- monotonically decreasing at all three levels tested, unlike the thin-slab case above -- toward the Fresnel value $R_F=0.0337$ of Table~\ref{tab:params}; the residual \rev{difference of about $0.3\%$ (within the rounding of the quoted values)} at the finest level tested is consistent with the small, physical (rather than numerical) broadening expected for a finite-bandwidth pulse reflecting off a weakly dispersive interface, superposed on a small remaining discretization error.

\subsection{Discussion}

The agreement between the numerically extracted front-surface reflectivity and the Fresnel formula, together with the correct timing of the back-surface echo and the constructive-interference enhancement found in the thin-slab regime, supports the internal consistency of the local (ADE-FDTD) reformulation of the integral equations \eqref{eq:Eintegral} and \eqref{eq:Pdotsol}, and, by the equivalence argument of Sec.~\ref{sec:equivalence}, of the integral-equation formulation itself. The thin-slab result in particular shows that the same microscopic multiple-scattering mechanism that, in the continuous-wave limit, produces the familiar Airy fringes of thin-film optics, also correctly reproduces -- now resolved pulse-by-pulse in time -- the coherent, constructive or destructive, addition of a \emph{finite} number of reflected wavelets for a genuinely broadband excitation, a regime in which the usual infinite-geometric-series derivation of the Airy formula does not directly apply.

\section{Conclusions}
\label{sec:conclusions}

We have reformulated the propagation of a few-cycle light pulse through a thin dielectric slab as a pair of coupled, retarded integral equations, derived from a microscopic picture of the slab as a collection of polarizable atoms coupled by causally-retarded dipole radiation. We have shown that this integral formulation is exactly equivalent to a local system consisting of the wave equation for the field and a driven, damped harmonic-oscillator equation for the atomic polarization, and we have solved this local system numerically with a leapfrog finite-difference time-domain scheme, an auxiliary-differential-equation treatment of the Lorentz polarization, and a total-field/scattered-field source injection designed to respect, rather than override, the causal initial condition that only the incident field can be present before any reflection has had time to form.

For a slab thick compared to the pulse length, the numerically extracted front-surface reflectivity matches the Fresnel formula to within about $1\%$, and a distinct back-surface echo appears at exactly the expected round-trip delay: a direct time-domain confirmation of the multiple-scattering structure built into the theory. For a slab thinner than one optical cycle, the front- and back-surface reflections instead overlap and interfere, reproducing -- now resolved in time rather than swept in frequency -- the same coherent addition mechanism responsible for the classical thin-film (Airy) interference fringes; a space-time map of the field makes this interference directly visible as a cusp-shaped overlap region straddling the front interface.

Several extensions of this framework seem particularly promising for future work:
\begin{itemize}
    \item Realistic Material Dispersion: Replacing the present single-Lorentz illustrative parameters with multi-oscillator fits (such as a full Sellmeier expansion) to accurately match the dispersion and absorption profiles of real optical glasses over broad spectral ranges.
    \item Spatially Inhomogeneous Media: Extending the spatial grid to accommodate non-uniform atomic density profiles $N(z)$ or graded-index interfaces, allowing the study of pulse dynamics in engineered thin-film stacks, optical coatings, and photonic structures.
    \item Multidimensional and Oblique Incidence: Upgrading the 1D scalar solver to a full multidimensional electromagnetic code to investigate non-plane-wave focused beams and oblique incidence angles, where polarization-dependent phase shifts and spatial beam reshaping occur upon reflection and transmission.
    \item Parametric Crossover Sweeps: Systematically sweeping either the slab thickness at a fixed pulse duration or the pulse length at a fixed thickness to map the continuous quantitative crossover from the well-separated pulse-echo regime to the fully interfering time-domain Airy regime.
\end{itemize}

\rev{The equivalence between the local and the retarded formulations, proven analytically in Sec.~\ref{sec:equivalence}, is tested numerically in Paper~II~\cite{Reali2026II}: there the integral equations are solved directly in the time domain, and the reflected and transmitted fields, the energy balance and the full space-time field agree with the FDTD--ADE results reported here.}

As stressed in Sec.~\ref{sec:parameters}, the material parameters used here serve as an illustrative benchmark; applying this quantitative framework --- now fully validated and grid-converged (Sec.~\ref{sec:convergence}) --- to measured dispersion profiles of optical glasses and engineered structures represents the natural step forward in time-domain thin-film optics.

\section*{Acknowledgments}
The author wishes to thank his former student, former colleague, and dear friend Antoniangelo Agnesi for fruitful discussions and suggestions during the preparation of this work. \rev{The author also acknowledges the use of Claude (Anthropic), an AI assistant, for help with the numerical analysis and the editorial revision; the author takes full responsibility for the content.}

\appendix
\section{Energy conservation in the Lorentz model}
\label{app:energy}

We derive the energy-conservation relation $R+T+A=1$ used in Sec.~\ref{sec:results} (Eq.~\eqref{eq:energy}) to validate the numerical scheme, and give the exact discrete sums used to evaluate $R$, $T$, $A$ from the FDTD output.

\subsection{Poynting's theorem for the Maxwell--Lorentz system}

Introduce, purely for this derivation, the magnetic field $H(z,t)$ satisfying the standard 1D Maxwell curl equations equivalent to the wave equation \eqref{eq:wave} used throughout the paper,
\begin{equation}
\frac{\partial E}{\partial z} = -\frac{1}{c}\frac{\partial H}{\partial t}, \qquad
\frac{\partial H}{\partial z} = -\frac{1}{c}\left(\frac{\partial E}{\partial t}+4\pi\frac{\partial P}{\partial t}\right).
\label{eq:maxwell_pair}
\end{equation}
(Eliminating $H$ between the two recovers Eq.~\eqref{eq:wave} exactly; $H$ never appears elsewhere in the paper and is introduced here only to invoke Poynting's theorem.) With the electromagnetic energy density and flux
\begin{equation}
u_{EM} = \frac{E^2+H^2}{8\pi}, \qquad S = \frac{c}{4\pi}EH,
\end{equation}
a direct calculation using \eqref{eq:maxwell_pair} gives Poynting's theorem for this system,
\begin{equation}
\frac{\partial u_{EM}}{\partial t} + \frac{\partial S}{\partial z} = -E\,\frac{\partial P}{\partial t}.
\label{eq:poynting}
\end{equation}
The right-hand side is the rate of work done by the field on the polarization current; it is not yet the dissipated power, since the oscillator can also store energy reversibly.

\subsection{Oscillator energy and dissipation}

Write the Lorentz oscillator equation \eqref{eq:oscillator} as $\ddot P+2\gamma\dot P+\omega_0^2P=\kappa E$, with $\kappa\equiv Ne^2/m$. Dividing by $\kappa$ and multiplying by $\dot P$ gives
\begin{equation}
\frac{d}{dt}\left[\frac{1}{2\kappa}\Big(\dot P^2+\omega_0^2P^2\Big)\right] = E\dot P - \frac{2\gamma}{\kappa}\dot P^2.
\label{eq:oscenergy}
\end{equation}
The left-hand side is the time derivative of the oscillator's kinetic-plus-potential energy density,
\begin{equation}
u_{\mathrm{osc}} \equiv \frac{1}{2\kappa}\Big(\dot P^2+\omega_0^2P^2\Big),
\end{equation}
which has the same units as $u_{EM}$ (both being $\propto E^2$, since $\kappa$ has units of $\mathrm{s}^{-2}$ and $P$ the same units as $E$). The last term, $p_{\mathrm{abs}}(z,t)\equiv(2\gamma/\kappa)\dot P^2\ge0$, is therefore the genuinely dissipated power density, correctly normalized to match $\partial u_{EM}/\partial t$ in Eq.~\eqref{eq:poynting}. Combining \eqref{eq:poynting} and \eqref{eq:oscenergy} to eliminate $E\dot P$,
\begin{equation}
\frac{\partial}{\partial t}\big(u_{EM}+u_{\mathrm{osc}}\big) + \frac{\partial S}{\partial z} = -\frac{2\gamma}{\kappa}\left(\frac{\partial P}{\partial t}\right)^2 \le 0,
\label{eq:totalbalance}
\end{equation}
an exact, local conservation law: the total (field $+$ oscillator) energy density is conserved except for the dissipative sink on the right, which is nonzero only inside the slab (where $\kappa\neq0$).

\subsection{Integrated balance: $R+T+A=1$}

Integrate \eqref{eq:totalbalance} over $z\in[0,L]$ and $t\in[t_1,t_2]$, with $t_1$ before the pulse reaches the front interface and $t_2$ late enough that $u_{EM}+u_{\mathrm{osc}}$ has returned to (numerically) zero throughout $[0,L]$. The volume integral of $\partial(u_{EM}+u_{\mathrm{osc}})/\partial t$ then vanishes, leaving a pure boundary-flux balance,
\begin{equation}
\int_{t_1}^{t_2}\!\big[S(L,t)-S(0,t)\big]\,dt = -\int_{t_1}^{t_2}\!\!\int_0^L \frac{2\gamma}{\kappa}\left(\frac{\partial P}{\partial t}\right)^2 dz\,dt.
\label{eq:boundaryflux}
\end{equation}
At $z=L$ only the transmitted, right-going wave is present, so $H(L,t)=E_t(L,t)$ and $S(L,t)=(c/4\pi)E_t(L,t)^2$. At $z=0$ the total field is the superposition $E(0,t)=E_{in}(0,t)+E_r(0,t)$ of the incident (right-going, $H=+E_{in}$) and reflected (left-going, $H=-E_r$) waves, so $H(0,t)=E_{in}(0,t)-E_r(0,t)$ and
\begin{equation}
S(0,t) = \frac{c}{4\pi}\big[E_{in}(0,t)+E_r(0,t)\big]\big[E_{in}(0,t)-E_r(0,t)\big] = \frac{c}{4\pi}\Big[E_{in}(0,t)^2-E_r(0,t)^2\Big].
\end{equation}
Substituting both into \eqref{eq:boundaryflux} and rearranging,
\begin{equation}
\underbrace{\frac{c}{4\pi}\int E_{in}^2\,dt}_{E_{\mathrm{inc}}} =
\underbrace{\frac{c}{4\pi}\int E_r^2\,dt}_{E_{\mathrm{ref}}} +
\underbrace{\frac{c}{4\pi}\int E_t^2\,dt}_{E_{\mathrm{trans}}} +
\underbrace{\int\!\!\int_0^L \frac{2\gamma}{\kappa}\left(\frac{\partial P}{\partial t}\right)^2 dz\,dt}_{E_{\mathrm{abs}}},
\label{eq:energybalance}
\end{equation}
i.e.\ $E_{\mathrm{inc}}=E_{\mathrm{ref}}+E_{\mathrm{trans}}+E_{\mathrm{abs}}$. Dividing by $E_{\mathrm{inc}}$ gives the identity used in the main text,
\begin{equation}
R+T+A=1, \qquad R\equiv\frac{E_{\mathrm{ref}}}{E_{\mathrm{inc}}},\quad T\equiv\frac{E_{\mathrm{trans}}}{E_{\mathrm{inc}}},\quad A\equiv\frac{E_{\mathrm{abs}}}{E_{\mathrm{inc}}}.
\end{equation}

\subsection{Discrete evaluation}

In the FDTD implementation, the three energies in \eqref{eq:energybalance} are evaluated as the discrete sums
\begin{equation}
E_{\mathrm{inc}} = \frac{c}{4\pi}\sum_n E_{in}^2(t_n)\,\Delta t, \qquad
E_{\mathrm{ref}} = \frac{c}{4\pi}\sum_n E_r^2(t_n)\,\Delta t, \qquad
E_{\mathrm{trans}} = \frac{c}{4\pi}\sum_n E_t^2(t_n)\,\Delta t,
\label{eq:discretesums1}
\end{equation}
\begin{equation}
E_{\mathrm{abs}} = \frac{2\gamma}{\kappa}\sum_n\Delta t\sum_i \Delta z\,\dot P_i^2(t_n),
\label{eq:discretesums2}
\end{equation}
where $E_{in}$, $E_r$, $E_t$ are read off the probes described in Sec.~\ref{sec:numerics}, and $\dot P_i(t_n)$ is the centered-difference estimate $[P_i^{n+1}-P_i^{n-1}]/(2\Delta t)$ accumulated at every grid point inside the slab and at every time step.

Evaluating \eqref{eq:discretesums1}--\eqref{eq:discretesums2} for the parameters of Table~\ref{tab:params} gives, exactly as quoted in Sec.~\ref{sec:results}, \rev{$R=0.0626$, $T=0.8826$, $A=0.0542$ (sum $0.9994$)} for the thick slab, and $R=0.120$, $T=0.878$, $A=0.003$ (sum $1.000$) for the thin slab. The $\lesssim0.1\%$ deviation of $R+T+A$ from unity in both cases is consistent with the residual discretization error of the second-order leapfrog scheme quantified in Sec.~\ref{sec:convergence}, and the larger absorbed fraction for the thick slab simply reflects the longer optical path traveled inside the (weakly) lossy medium. Beyond validating the numerical scheme, this identity is a concrete illustration, for students, of how a global physical conservation law -- here following directly from Poynting's theorem applied to the Maxwell--Lorentz system -- can be turned into an independent, quantitative check on a computational result.

\section{Glossary of numerical-methods terms}
\label{app:glossary}

For readers less familiar with finite-difference time-domain methods, we briefly define here three terms used throughout Secs.~\ref{sec:numerics}--\ref{sec:convergence}.

\paragraph{Courant stability limit.}
An explicit leapfrog scheme for the wave equation \eqref{eq:wave}, such as Eq.~\eqref{eq:leapfrog}, is stable only if numerical information cannot propagate faster than one grid cell $\Delta z$ per time step $\Delta t$ -- i.e.\ no faster than the physical speed of light $c$ allows. This requires the Courant number
\begin{equation}
S \equiv \frac{c\,\Delta t}{\Delta z} \le 1
\end{equation}
in one spatial dimension; violating it makes numerical errors grow without bound instead of propagating and canceling correctly. Throughout this paper we use $S=0.4$, comfortably inside the stable range.

\paragraph{Richardson (empirical) order of convergence.}
If a numerical scheme is formally accurate to order $p$, the error should shrink as $\Delta z^p$ as the grid is refined. Given a diagnostic $f(\Delta z)$ computed at three grid spacings successively halved, $\Delta z$, $\Delta z/2$, $\Delta z/4$, the \emph{observed} order can be estimated, without knowing the exact solution, from the ratio of successive differences,
\begin{equation}
p \approx \log_2\!\left|\frac{f(\Delta z)-f(\Delta z/2)}{f(\Delta z/2)-f(\Delta z/4)}\right|.
\end{equation}
This is the quantity reported in Sec.~\ref{sec:convergence} ($p\approx2.0$), confirming the expected second-order accuracy of the leapfrog scheme; the same three values also allow a Richardson \emph{extrapolation} to a more accurate estimate of the (unknown) continuum limit.

\paragraph{Mur absorbing boundary condition.}
A finite computational domain needs boundaries that let outgoing waves leave without reflecting back in, mimicking open (infinite) space. Mur's first-order condition \cite{Mur} enforces this locally, at each boundary node, by discretizing a one-way wave equation that admits only outgoing solutions; for the right boundary (node $N$) used in Eq.~\eqref{eq:leapfrog}, it reads
\begin{equation}
E_N^{n+1} = E_{N-1}^{n} + \frac{c\Delta t-\Delta z}{c\Delta t+\Delta z}\Big(E_{N-1}^{n+1}-E_N^{n}\Big),
\end{equation}
with the mirror-image expression at the left boundary. It is exact only for waves striking the boundary at normal incidence -- sufficient here, since the geometry is strictly one-dimensional.

\end{document}